# Robustness of Quantum Random Walk Search Algorithm in Hypercube when only first or both first and second neighbors are measured

*Hristo Tonchev[1], Petar Danev[1]*

[1] *Institute for Nuclear Research and Nuclear Energy, Bulgarian Academy of Sciences,72 Tzarigradsko Chaussée, 1784 Sofia, Bulgaria*
Emails: htonchev@issp.bas.bg        pdanev@inrne.bas.bg

**Abstract:** *In this work we study the robustness of two modifications of quantum random walk search algorithm on hypercube. In the first previously suggested modification, on each even iteration only quantum walk is applied. And in the second, the closest neighbors of the solution are measured classically. In our approach the traversing coin is constructed by both generalized Householder reflection and an additional phase multiplier and we investigate the stability of the algorithm to deviations in those phases. We have shown that the unmodified algorithm becomes more robust when a certain relation between those phases is preserved. The first modification we study here does not lead to any change in the robustness of quantum random walk search algorithm. However, when a measurement of the first and second neighbors is included, there are some differences. The most important one, in view of our study of the robustness, is an increase in the stability of the algorithm, especially for large coin dimensions.*

**Keywords:** *Quantum information, Quantum algorithms, Quantum Random Walk, Quantum Search, Generalized Householder Reflection*

## 1. Introduction:

Discrete time Quantum random walk search algorithm (QRWS) was first proposed by Shenvi *et al.* in their work [1]. It is probabilistic algorithm, based on quantum walk [2]. Together with the more famous Grover's search algorithm [3] QRWS can find the searched element quadratically faster than the best-known classical algorithm. However, unlike the Grover's algorithm that can be used only on linear database, the QRWS can be used on structures with arbitrary topologies. Search on structures of particular interest are simplex [4], square and cubic grids [5], tree graph [6], hypercube [7] and fractal structures [8]. The ability of the quantum walks to quickly find an element on particular structure makes the algorithm a very important component in various other quantum algorithms. Some examples are: the quantum algorithm for graph's isomorphism [9], the algorithm for calculating Boolean formulas [10], and the algorithm for finding subgraphs with particular shape in a larger graph [11].

There are different modifications of quantum walk search, that lead to improvements for a particular structure. Like for example in [7], two modifications for hypercube were proposed: the first is to apply only quantum walk on each even

iteration – this change doesn't affect the probability to find a solution but decreases the number of iterations needed to find the solution. The second modification adds a classical measurement of the closest neighbors of the state returned by the algorithm. This modification doubles the probability to find a solution. Another recent works [12][13], study QRWS when there is more than one solution.

In order to implement any quantum algorithm, there should be an efficient way to construct quantum gates. These methods are highly dependent on the physical system used to perform the quantum computation. Two popular such methods that can be used for constructing both quantum gates for qubits and qudits are – by using Givens rotations [3] and by decomposition to generalized Householder reflections [14]. The second method is quadratically faster, and can be applied on physical systems like ion trap [15] and photonic quantum computers [16].

In our previous works [17] [18] we study quantum random walk search algorithm with traversing coin constructed by using generalized Householder reflection and a phase multiplier. Such walk coin has two phases - one in the multiplier and one in the Householder reflection. We have shown that if a particular functional dependence between them is maintained, the algorithm become more robust – probability to find solution remains high even if there is an inaccuracy in the phases. However, as the size of the coin increases, the time and memory required by the calculation are proportional to the power of two of the coin size. That is why in those works we use deep neural network to make prediction for larger coin size. In our previous work [19], we used a logistic regression to make the predictions - a slightly modified Hill functions to approximate the probability to find solution as a function of the Householder reflection's phase. Next, we extrapolate the results for larger coin size.

The Hill function [20] is non-linear logistic regression function that have a variety of applications in mathematical and computational biology. Its form is similar to the one of sigmoid function but the slope of the curve could be controlled. This function could not be mistaken with the Hill equations in quantum mechanics [21], that have the same name but entirely different meaning – a solution of a second order differential equation for a periodic function.

This work's main goal is to use numerical calculations and prognoses based on logistic regressions with Hill function to show that measuring first and second neighbors increases the robustness of the quantum random walk search algorithm on hypercube with walk coin construct by generalized Householder reflection and a phase multiplier.

This article is organized as follows: in Sec. [2] we give a brief description of the QRWS algorithm and two of its well-known modifications. In Subsec. [2.1] we describe the original algorithm without optimization to a particular graph structure. In the next Subsec. [2.2], one modification for hypercube is discussed - replacing half of algorithm's iterations with quantum walk only. Subsec. [2.3] reviews another modification of QRWS algorithm that leads to an increase of the probability to find the solution when the nearest neighbours to the state found by the algorithm are measured too. In Sec. [3] we recap a modification for hypercube that increases the algorithm's robustness. In Subsec. [3.1] the construction of the walk coin with generalized Householder reflection and a phase multiplier is discussed. The dependence of the probability to find a solution on these phases is evaluated. In Subsec. [3.2] we study the algorithm's success rate when a particular functional dependence between the phases is introduced. Subsec. [3.3] defines the robustness of the algorithm, and in Subsec. [3.4] we show the modification of the Hill function used to fit the probability to find solution for some functional dependences between the coin phases. In Sec. [4] and Sec. [5] are our results in this work. In Sec. [4] we discuss that modifying the algorithm to use quantum walk on even iterations leads to the same robustness as the unmodified algorithm. Sec. [5] shows our result for QRWS with our coin and with additional measurements of the first and second neighbors. Subsec. [5.1] discusses the probability to find solutions for specific functional dependences between phases, when only first or first and second neighbors' probabilities are taken into account. Next, in Subsec. [5.2] we show our numerical results for approximations of QRWS parameters with a Hill function. In Subsection [5.3] we discuss the robustness of the quantum algorithm for some functional dependencies between the coin phases and the improvement obtained by including the first or both the first and second neighbors. We give prognosis for larger coin sizes based on Hill functions extrapolations in [5.4]. In Sec. [6] we summarize our results.

## 2. Quantum random walk search on hypercube
### 2.1. General description of the QRWS on an arbitrary structure

Quantum random walk search algorithm was first introduced in [1]. It is based on quantum walk - quantum analogue of the classical random walk. Both quantum random walk search and Grover search are algorithms that search in unordered database quadratically faster than a classical algorithm. In contrast to Grover's algorithm that can be used to search only a linear database, the quantum random walk search algorithm can be used to search for an element in the graph with arbitrary topology. On Fig. 1 is shown the quantum circuit of QRWS algorithm:

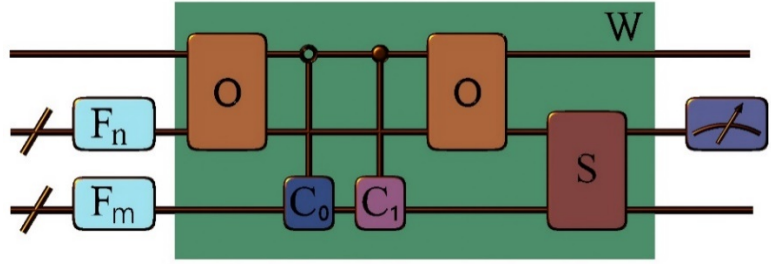

Fig. 1. *The quantum circuit of QRWS algorithm. An iteration of the algorithm is denoted by W. The quantum gates that take part in the iteration are: oracle O, discrete Fourier transformation with appropriate dimensions $F_n$ and $F_m$, Shift operator S, walking and marking coins $C_0$ and $C_1$.*

The algorithm uses three registers – a control register consisting of one qubit, a node register containing $n$ states, and a coin register (known also as an edge register) with m states. In case of Hypercube $n = 2^m$. The size of the register will be denoted with a subscript after the ket vector. The initial state of the algorithm's register is:

$$|\Psi_0\rangle = |0\rangle_2 \otimes |0\rangle_n \otimes |0\rangle_m = |0,0,0\rangle_{2 \times m \times n} \qquad (1)$$

The algorithm begins by putting the node and coin registers in an equal weight superposition. This can be done by applying Discrete Fourier Transformation (DFT) ($F_n$ and $F_m$) operators on those registers. The control register's state remains $|0\rangle_2$.

On the whole algorithm's register, the QRWS iteration $W$ should be applied $k$ times. The iteration consists of the following sequence of operators:

1) First, an oracle (O) is applied on control and node registers, entangling them. It is able to recognize if the given node is a solution or not. It marks the solutions by changing the state of the control register. The oracle can mark multiple states.

$$\mathbb{O}|q,i,j\rangle_{2 \times m \times n} = (O \otimes I_m)|q,i,j\rangle_{2 \times m \times n} \qquad (2)$$
$$= \begin{cases} |(q+1) \bmod 2, i, j\rangle_{2 \times m \times n} & j \in \{h_1, h_2, \ldots, h_\lambda\} \\ |q,i,j\rangle_{2 \times m \times n} & otherwise \end{cases}$$

2) On all states that are not solution is applied a traversing coin $C_0$. The coin can be any unitary matrix with dimension of the node register, however the one that gives the best result depends on the topology and dimension of the walked structure. In the case of a hypercube, the most commonly used walking coin is the Grover coin G.

$$C_0 = G = \hat{I}_m - 2|\chi\rangle\langle\chi| \tag{3}$$

$$\mathbb{C}_0 = \begin{pmatrix} \hat{I}_n \otimes C_0 & \hat{0}_{m \times n} \\ \hat{0}_{m \times n} & \hat{I}_{m \times n} \end{pmatrix} = Diag(\hat{I}_n \otimes C_0, \hat{I}_{m \times n},) \tag{4}$$

Where $\hat{I}_n$ is the identity matrix with size $n$, $|\chi\rangle$ is equal weight superposition, and $\hat{0}_{m \times n}$ are matrices with the same dimension filled with zeros.

3) The marking coin $C_1$ is applied on the states that oracle recognizes as solutions. The best marking coin depends again on the chosen topology and of the traversing coin. In the case of hypercube, the most common marking coin is a minus identity operator.

$$C_1 = -\hat{I}_m \tag{5}$$

$$\mathbb{C}_1 = \begin{pmatrix} \hat{I}_{m \times n} & \hat{0}_{m \times n} \\ \hat{0}_{m \times n} & \hat{I}_n \otimes C_1 \end{pmatrix} = Diag(\hat{I}_{m \times n}, \hat{I}_n \otimes C_1) \tag{6}$$

4) A second application of the oracle is made to the control and node registers, disentangling them.
5) The iteration ends by applying the Shift operator $S$ on the node and edge registers. Depending on the state of the edge registers, it executes the quantum walk. The shift operator determines nodes connected by an edge, and in this way defines the topology of the walked structure.

$$\mathbb{S}|0, j, k\rangle = (\hat{I}_2 \otimes S)|0, j, k\rangle = |0, j, g(k, j)\rangle \tag{7}$$

To summarize, the iteration can be written as:

$$W = \mathbb{S} \cdot \mathbb{O} \cdot \mathbb{C}_0 \cdot \mathbb{C}_1 \cdot \mathbb{O} \tag{8}$$

The state of the algorithm's register after each iteration can be written recursively:

$$|\Psi_{i+1}\rangle = W|\Psi_i\rangle \tag{9}$$

In case of one solution, the algorithm requires $k$ iterations, each of them containing two oracle calls.

$$k = \left\lceil \frac{\pi}{2}\sqrt{n/2} \right\rceil = \left\lceil \frac{\pi}{2}\sqrt{2^{m-1}} \right\rceil \tag{10}$$

Where [m] denotes the rounded-up m.

After the required number of iterations are made, the node register should be measured. The probability to find solution is approximately:

$$p = 1/2 - \mathcal{O}(1/2^m) \quad (11)$$

If the solution is not found, the algorithm can be repeated.

On Fig 2 are shown results from simulation of QRWS for coin size 6. The probability to find solution at each iteration is presented on the left. The probability to obtain each of the nodes after $k$ iterations is shown on the right.

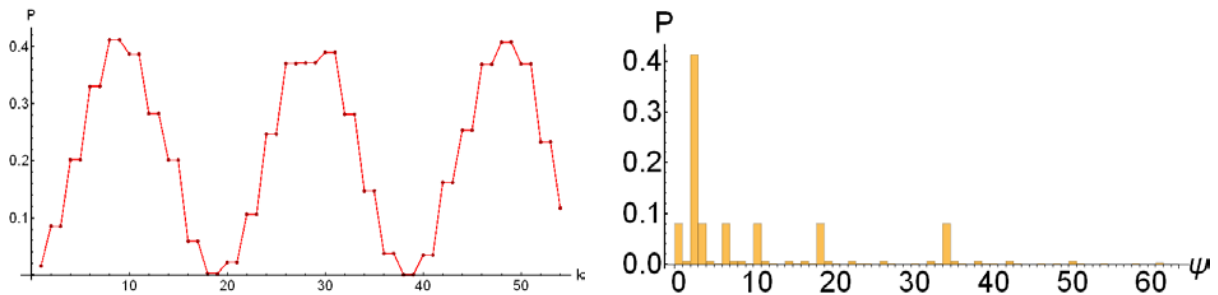

Fig. 2. *On the left picture is shown the probability to find solution of the QRWS algorithm with coin size 6 on each iteration between 1 and 56. The probability find the solution on iteration 2i and 2i+1 is the same. On the right is shown the probability to obtain each state after implementing QRWS with coin size 6, marked element $|3\rangle$ and k=9 iterations.*

In case of hypercube there exists different modifications that improves the algorithm. For example, Potocek *et al.* in their work [7] introduce two optimizations by using the properties of the random walk and the hypercube's symmetries:

1) The first one consists of combining QRWS with more classical measurements of the neighbours of the result obtained after applying the algorithm. This modification greatly increases the probability of finding a solution, at the expense of increasing the oracle calls.
2) The second modification combines QRWS with just QRW. This improvement halves the quantity of the oracle calls needed to implement the quantum random walk search algorithm.

Those modifications will be briefly explained in the next two sections.

## 2.2. Improving the quantum circuit by applying an oracle only on the odd steps

The quantum walk on hypercube, can be projected onto quantum walk on a line. The nodes with the same Hamming weight are projected onto the same point on the line (Fig. 3). If the quantum walker is on even positions during the $i$-th step, on the $i + 1$-th it will be on odd positions. In this way the quantum walk can be split into two independent quantum walks on a line –the first on the even elements and the second on the odd elements.

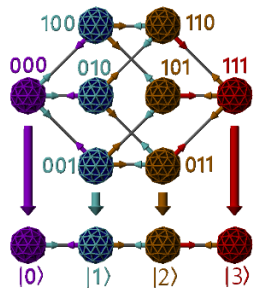

*Fig. 3. Projection of the quantum random walk on a hypercube onto quantum random walk on a line.*

This property allows additional optimization of QRWS on hypercube. It uses two different versions of QRWS iteration: on all odd iterations the standard QRWS iteration (the same as in Eq. 8) is implemented and on each even integration – only QRW on hypercube. The iteration operators can be written as follows:

$$W_{2i+1} = \mathbb{S} \cdot \mathbb{O} \cdot \mathbb{C}_0 \cdot \mathbb{C}_1 \cdot \mathbb{O} \qquad i = \{0,1,\dots\} \qquad (12)$$

$$W_{2i} = \hat{I}_2 \otimes \hat{I}_n \otimes C_0 \qquad i = \{0,1,\dots\} \qquad (13)$$

$$|\Psi_{j+1}\rangle = W_j |\Psi_j\rangle \qquad (14)$$

In this case, during the QRWS, only on the odd iterations the probability to find solution differs from the initial probability. An example for coin size 6 is shown on Fig. 4 left. This optimization only reduces the number of oracle calls needed in implementation of QRWS algorithm and didn't change the probability to find solution as can be seen in Fig. 4 right.

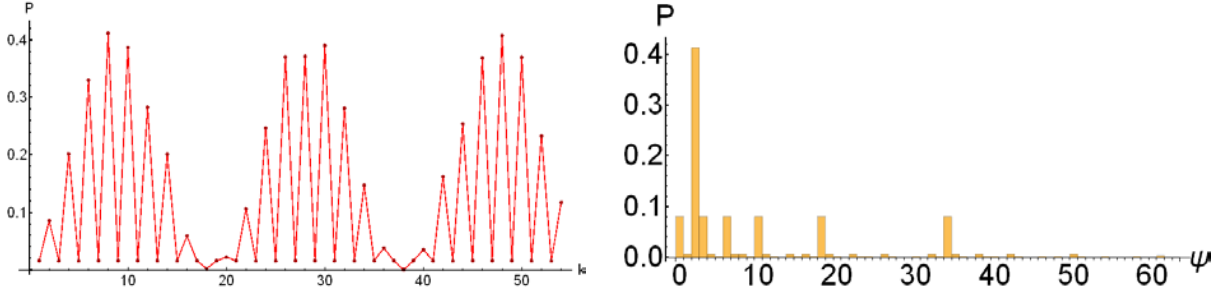

Fig. 4. On the left is shown the probability to find solution of QRWS algorithm when the oracle and marking coins are applied only on odd iterations. A simulation is made with coin size 6 for iterations between 1 and 56. On the right is shown the measured state after the required number of iterations. It coincides with the one of the standard QRWS on hypercube.

### 2.3. Classical measurement of the nearest neighbors

At the end of quantum random walk search algorithm, the state with the highest probability to be measured is the marked state. Its first neighbours have also relative high probability. If the marked state and its first neighbours are excluded, the highest probability have the second ones, next are the third and so on. Next neighbour differs by just one bit in the binary representation of the state marked as solution. Second by flipping two different bits and so on. Number of $o$-th neighbours in a hypercube with coin size m can be calculated as:

$$l(m, o) = \frac{m!}{(m-o)!\, o!} \qquad (15)$$

This modification relies on measuring not only the state that the algorithm returns after its execution, but also its neighbours. Measurement of its first neighbours slightly more than doubles the probability to find solution at the price of just m additional measurements (Eqs. (16) and (17)). The second neighbours do not give such a large increase of the probability to find solution. They are also too many, so in most of the cases it is not reasonable to measure them and other higher-order neighbours.

$$k = \left\lceil \frac{\pi}{2}\sqrt{n/2} \right\rceil + m = \left\lceil \frac{\pi}{2}\sqrt{2^{m-1}} \right\rceil + m \qquad (16)$$

$$p = 1 - \mathcal{O}(1/2^m) \qquad (17)$$

As an example, in the case of coin size $m = 6$ and we have in total 64 nodes. The marked state is $|2\rangle$ (see right sides of Fig 2 and Fig. 4). Here we have six first neighbours - $|0\rangle$, $|6\rangle$, $|10\rangle$, $|3\rangle$, $|18\rangle$ and $|34\rangle$. The second neighbours are 15 and so on. In this case, the probability of the unmodified quantum random walk search to succeeds and the contribution of the nearest neighbours are given in Table 1.

| Neighbour | None | First | Second | Residue |
|---|---|---|---|---|
| Probability to those states | 0.411765 | 0.479585 | 0.0814671 | 0.0271825 |
| Number of states | 1 | 6 | 15 | 42 |

*Table. 1. Probability to find solution in the case of coin size 6, depending on additional measurement of the nearest neighbours. The first row shows the consecutive neighbour, the second row gives the probability of finding a solution, which is added by including that neighbour, and the third row gives the number of additional measurements needed to achieve it.*

## 3. Walk coin modification
### 3.1. Construction of the walk coins by generalized Householder reflection

Different physical realizations of quantum computers rely on laws that governs the state and evolution of that particular physical system. That is why on some systems particular operations can be implemented much more efficiently.

One such example is the generalized Householder reflection, which can be done efficiently on some physical systems such as photonic and ion trap-based quantum computers. An easy way to construct a coin in these physical systems would be through a generalized Householder reflection with a phase multiplier [22]:

$$\mathbb{C}_0(\phi, \zeta, m) = e^{i\zeta}\left(\hat{I}_m - (1 - e^{i\phi})|\chi\rangle\langle\chi|\right) \quad (18)$$

The probability to find solution depends on the particular traversing and mark coins used. If in Eq. (18) $\phi = \zeta = \pi$, the Grover coin is obtained.

However, the physical systems are not ideal so during the realizations the gates cannot be constructed with perfect precision. For example, the inaccuracies in the laser frequency or phase, pulse shape and others. Noise can come also from interaction between the experimental setup and various external sources. This makes it very important to study the different types of noise and how they affect the quantum algorithm. This includes inaccuracies in the phases of the traversing [17] and mark [23] coins.

The probability to find solution of the unmodified QRWS as a function of phases for coin sizes 6 and 10 are shown on Fig. 5. It can be seen that the high probability areas are connected and form a stripe that depends on the coin size.

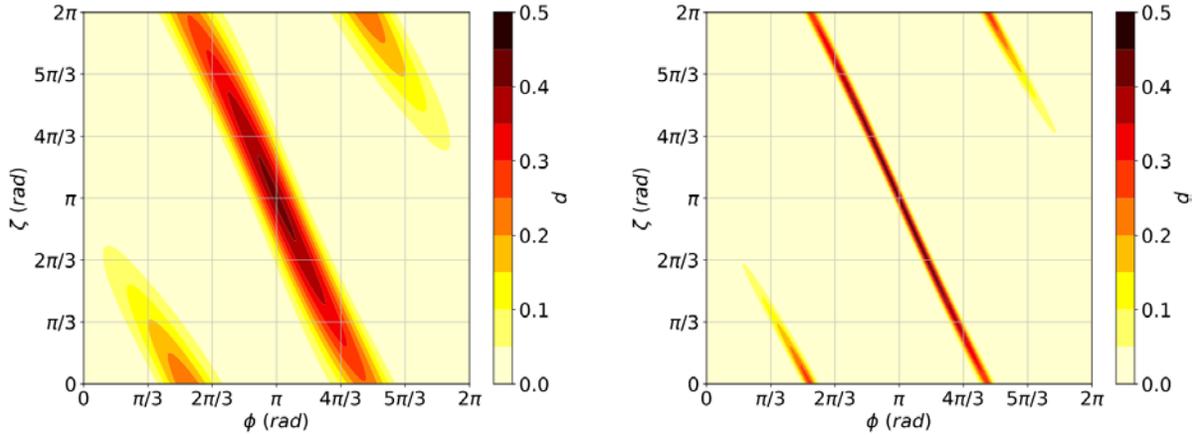

*Fig. 5. Probability to find a solution as a function of the generalized Householder phase ϕ and the additional phase multiplier ζ.*

There is a similar stripe with a high probability of finding a solution for all previously studied coin sizes between 4 and 11. The points with the highest probability P can be approximated with a functional dependence between the angles $\zeta(\phi)$:

$$P(\phi, \zeta, m) = P(\phi, \zeta(\phi), m = const) = P(\phi) \qquad (19)$$

## 3.2. Different functional dependences between phases

Here, we will show some examples of functional dependences between phases that will be used further in the text. Two of them are linear:

$$\zeta = \pi \qquad (20)$$

$$\zeta = -2\phi + 3\pi \qquad (21)$$

The first one shows the case when only the phase in the Householder reflection ϕ is controlled and ζ has a constant value. The second one corresponds to the line connecting two points in the space spawned by $\{\phi, \zeta\}$ with the highest probability - namely $\{\pi, -\pi\}$ and $\{-\pi, 5\pi\}$.

The other two have a small nonlinear correction:

$$\zeta = -2\phi + 3\pi - \left(\frac{1}{2\pi}\right)\sin(2\phi) \quad (22)$$

$$\zeta = -2\phi + 3\pi + \alpha_{ML}\sin(2\phi) \quad (23)$$

The first functional dependence is relatively close to the high probability stripe, and second one is found to give even better approximation. The parameter $\alpha_{ML}$ was derived by machine learning and its values for coin sizes between 4 and 11 can be found in [24].

Results from simulations of probability for finding a solution of QRWS with coin constructed according to Eq. (18) in case of coin sizes 6 and 10 are shown on Fig. (6). The different relations between the angles are depicted with different color and dashing. The red dot-dashed line, teal dashed and solid green and dotted blue correspond to Eq. (21), Eq. (20), Eq. (23), Eq. (22) respectively.

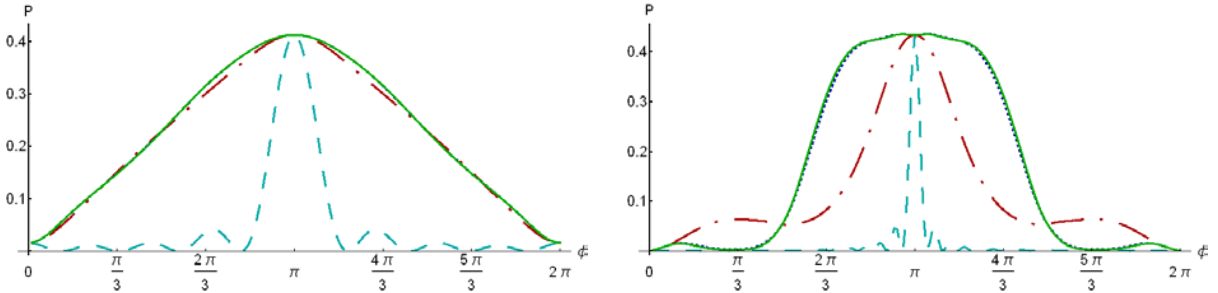

Fig. 6. *Probability to find solution depending on the phase in the generalized Householder reflection when there are different functional dependences between the walk coin phases. The dashed teal, red dash-dotted, dotted blue and solid green lines corresponds to Eq. (20), Eq. (21), Eq. (22), and Eq. (23).*

### 3.3. Robustness of the algorithm with our coin modification

In order to compare different functional dependences' stability against errors in the phase, we should introduce a way to measure it. We define the robustness $\varepsilon$ by the following:

$$p(\phi \in (\phi_{max} - \varepsilon^-, \phi_{max} + \varepsilon^+)) \cong p_{max} \equiv p(\phi_{max}) \quad (24)$$

Where $p_{max}$ is the maximal probability to find solution for a particular functional dependence between the angles $\zeta(\phi)$ and $\phi_{max}$ is the angle $\phi$ where this is achieved. The interval $\Delta = (\phi_{max} - \varepsilon, \phi_{max} + \varepsilon)$ is symmetric about the point $\phi_{max}$, so $\varepsilon^- = \varepsilon^+ = \varepsilon$.

As $\varepsilon$ increases the algorithm becomes more robust against inaccuracies in the phase. However, in many cases more practical approach is to define robustness at a percentage $\Omega$ of the maximal probability, not at the maximal probability itself:

$$p(\phi \in (\phi_{max} - \varepsilon, \phi_{max} + \varepsilon)) \geq \Omega p_{max} \qquad (25)$$

In this work, in the above equation we will use $\Omega = 0.9$.

On Fig. 7 is shown the robustness of QRWS algorithm for all four functional dependences mentioned before for different coin sizes. It can be seen that the least robust is the algorithm with coin parameters given by Eq. (20), the best linear dependence - by Eq. (21), the nonlinear dependence Eq. (22) gives much better results than the linear ones and the best results comes from Eq. (23).

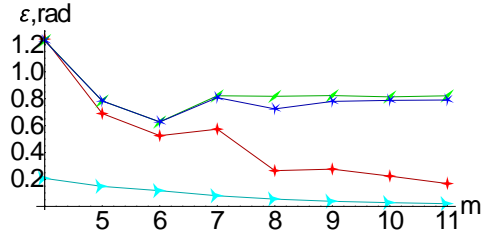

*Fig. 7. Robustness of QRWS algorithm for different coin sizes. Different colour and marker correspond to different coin parameter relations. The teal tree pointed star is used for Eq. (20), the red four pointed star corresponds to Eq. (21), the blue five pointed star - to Eq. (22), and the green two pointed star - to Eq. (23).*

The robustness can be used not only to quantitively compare different functional dependences, but also different modifications of the algorithm [23].

### 3.4. Hill function approximations

The probability to find solution as function of the phase $p(\phi)$ can be approximated by using a modification of Hill function as first introduced in [19]:

$$W(\phi, b, k, n) = \frac{bk^n}{|\phi - \pi|^n + k^n} \qquad (26)$$

The modified formula has 3 parameters: $b$ corresponds to the maximal height, $k$ is width of the plateau and $n$ is the slope of the probability curve (see for example Fig. 6). Those parameters depend on the relation $\zeta(\phi)$ and coin size $m$. The goodness of fit can be assessed by using standard deviation:

$$\sigma = \sqrt{\sum_{j=1}^{N} \frac{\left(W_j(\phi, b, k, n) - P_j(\phi, \zeta(\phi), m)\right)^2}{N - q}} \quad (27)$$

Here, $N$ is the number of fitting points, $q$ - the number of fitting parameters. $W_j$ and $P_j$ correspond to the probability to find solution taken from simulations and from the fit for the j-th point $\phi = \phi_j$.

An example for fitting $P(\phi, \zeta)$ for the four functional dependences Eqs. (20÷23) and coin size 6 can be seen on Fig (8). Here, a fit for Eq. (20) is on the top left side, Eq. (21) on the top right, Eq. (22) and Eq. (23) are on the bottom left and right respectively. The solid line is from the data and the dashed line shows the fit with the Hill function.

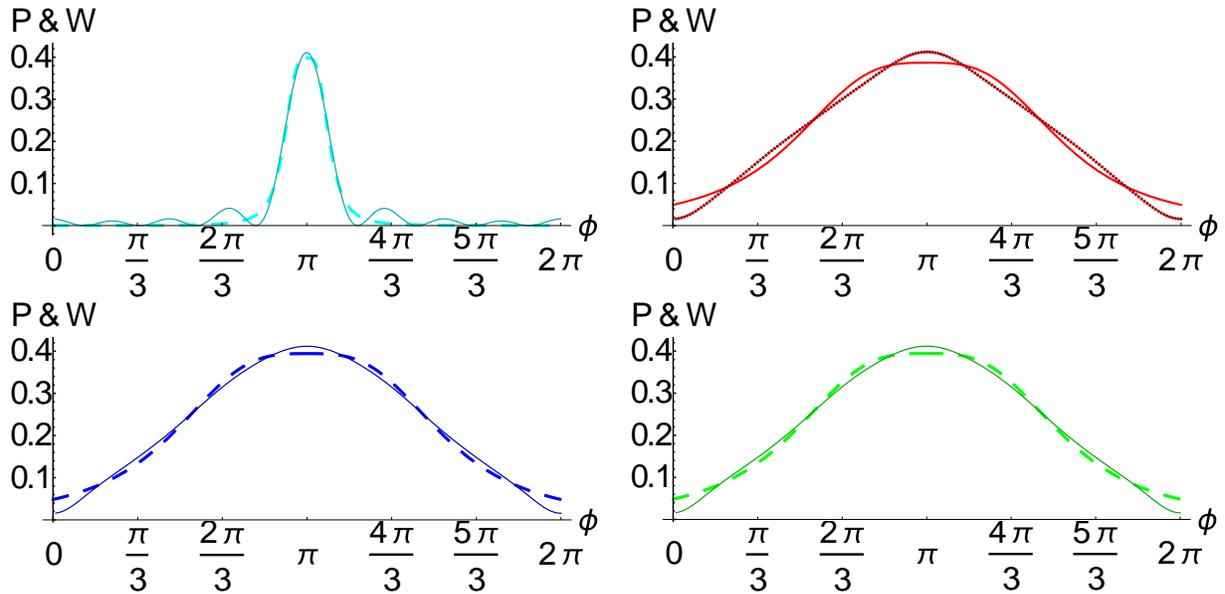

*Fig. 8. Comparison between Hill fit and simulation of QRWS when the coin size is 6. With solid dark lines are probability to find solution for functional dependence given by: Eq. (20) (top left), Eq. (21) (top right), Eq. (22) (bottom left) and Eq. (23) (bottom right). The dashed lines correspond to Hill fit for the same equation.*

The best fit parameters with the same functional dependence $\zeta_i(\phi)$, $i \in Eqs. (20 \div 23)$ but different coin size $m$, can be fitted themselves as a function of m:

$$b_i(m) = b(\zeta_i(\phi), m) \atop k_i(m) = k(\zeta_i(\phi), m) \atop n_i(m) = n(\zeta_i(\phi), m)\Bigg\} \; for\ the\ same\ dependence\ \zeta_i(\phi) \qquad (28)$$

Finally, the probability $W_j$ can be expressed as a function of only two parameters – the angle $\varphi$ and the coin size m:

$$W(\phi, m) = \frac{b_i(m) k_i(m)^{n_i(m)}}{|\phi - \pi|^{n_i(m)} + k_i(m)^{n_i(m)}} \qquad (29)$$

By using this approximation, a prognosis for different characteristic of the QRWS algorithm with our coin can be obtained. For example, in [19] we obtain lower bound for the maximal probability and the robustness of the algorithm by the Hill fit. This approximation can also be used to compare different properties for different functional dependence between phases and/or different modifications of the algorithm.

## 4. Robustness of the modifications
### 4.1. Modified circuit

This modification, reviewed in Sec. 2.2., compared to the standard QRWS algorithm, differs only in the case of odd iteration. If the state is measured on odd iteration the probability to find solution is the same as in the original algorithm. That leads to the same robustness as in the case of the unmodified algorithm.

### 4.2. Inclusion of nearest neighbors
#### 4.2.1. Probability to find solution and robustness for different functional dependences and number of neighbors

The probability to find solution when first neighbors of the solution are measured too, is slightly more than twice that of the standard QRWS on hypercube for the cost of m additional classical measurement. Here we will study how robust is this modification against inaccuracies in the phase $\phi$.

When only the first neighbors are included, probability to find solution more than doubles. Inclusion of the second neighbors increases the probability to find solution and slightly increase algorithm's robustness for additional $m(m-1)/2$ measurements. This result is numerically verified for all coin sizes between 4 and 11. For small coin size those additional measurements, that come from second neighbours, have diminishing returns. For example, in case of coin size 10 there is approximately 10 percent increasement in the probability. For larger coin size $m(m-1)/2 \ll 2^m$, it can be helpful to increase the robustness by adding second neighbors.

Results from simulations for the probability to find solution in case of different relations $\zeta(\phi)$ when the coin sizes 6 and 10 are shown on Fig 9 and for all coin sizes between 4 and 11 is shown in Appendix 1. Simulations for the probability to find solution of the unmodified QRWS $P_W(\phi)$ are shown on the first row, second row shows the probability when first neighbors are also taken $P_F(\phi)$, and when the first and second neighbors are included $P_S(\phi)$ the pictures on the third row are obtained. The red dot-dashed, dashed teal, solid green, and dotted blue lines corresponds to Eq. (21), Eq. (20), Eq. (23) and Eq. (22) respectively.

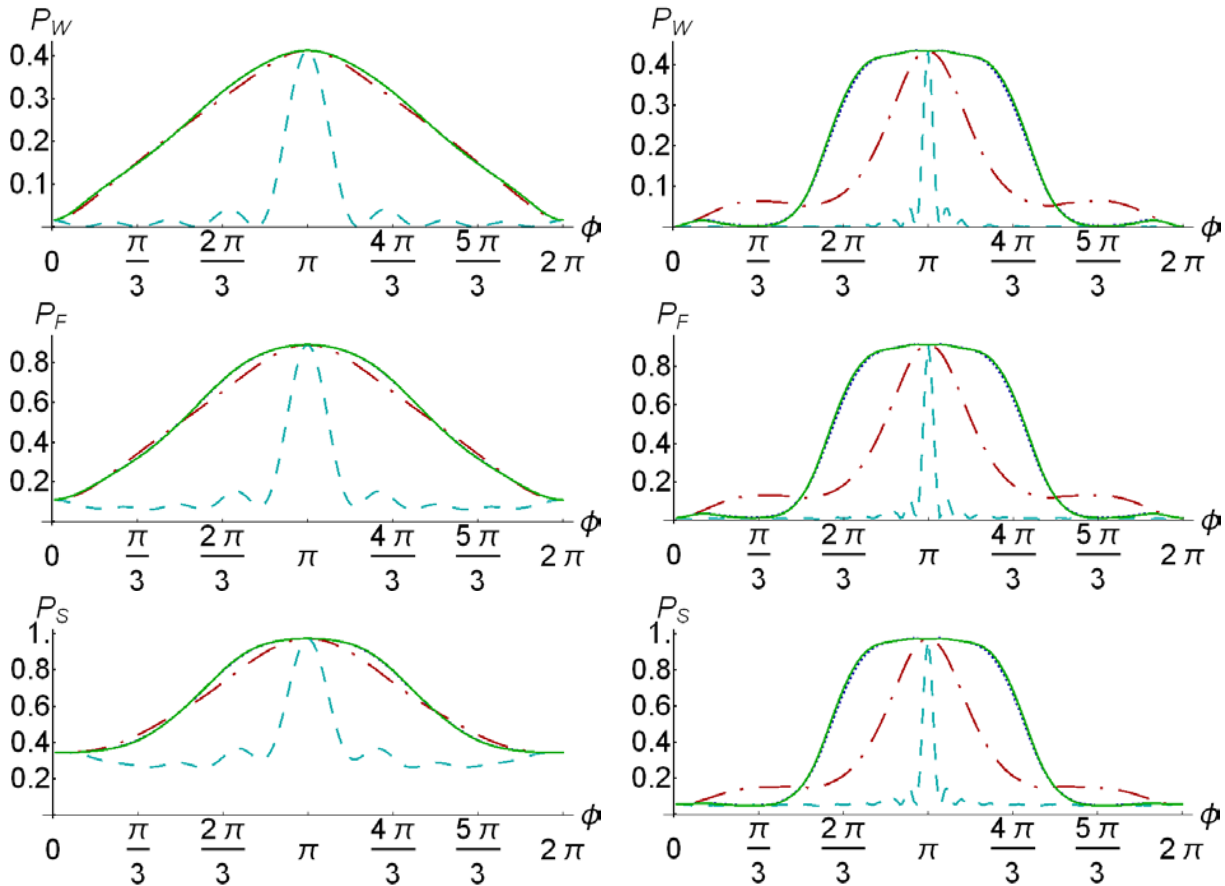

Fig. 9. Probability to find solution of QRWS algorithm as a function of the Householder phases for coin sizes 6 and 10. Different colour corresponds to different functional dependence between the coin phases. The dashed teal, red dash-dotted, dotted blue, and solid green lines corresponds to Eq. (20), Eq. (21), Eq. (22) and Eq. (23). The first row corresponds to the case when no neighbors are included. The second row includes first neighbours and the third row includes both first and second neighbours.

The dependence between the probability P and the phase $\phi$ of those modification have a very similar behavior to the unmodified QRWS on hypercube. The shape of the curve is similar and the highest probability areas are at the same values of $\phi$. However, there is one notable difference. The probability P close to $\phi = 0$ and $\phi = 2\pi$ is increased. This is due to the added probabilities from the first (or first and second) neighbors. When the coin is with $\phi \gtrsim 0$ (where $\gtrsim$ means larger but close to) or $\phi \lesssim 2\pi$, then the probability to find solution is approximately the same for all states of the node register. This leads to m times increase of P when only the first neighbors are taken and m(m-1)/2 times increase when both first and second neighbors are measured. The results for all other coin sizes in interval between 4 and 11 are similar.

Numerical simulations can be used only for relatively small coin sizes. In the next section we will use method for fitting with the Hill function to make an approximation of the $P(\phi, \zeta(\phi), \phi)$ for larger values of coin size $m$.

### 4.2.2. Approximation by Hill function

Similarly, to the case of the original QRWS algorithm, a modification that includes only first neighbors or both first and second neighbors can be approximated by using the Hill function.

Here we will make an approximation of the success probability of QRWS algorithm in both cases when only first neighbors are taken ($P_F(m, \zeta(\phi), \phi)$) and when both first and second neighbors are used ($P_S(m, \zeta(\phi), \phi)$). In case of Eq. (21), Eq. (22), Eq. (23), the whole interval of values in $\phi \in (0, 2\pi)$ is taken in order to make the approximation.

However, in the case of Eq. (20) there are negligible number of points in the central peak. So, to improve the approximation in this case of Eq. (20), we take only the points in a smaller interval around the high probability central peak $\phi \in (2\pi/3, 4\pi/3)$.

As an example, on Fig (10) we give the Hill approximation of $P_F(m, \zeta(\phi), \phi)$ for coin size 6 with dependence between phases according to Eq. (20), Eq. (21), Eq. (22) and Eq. (23).

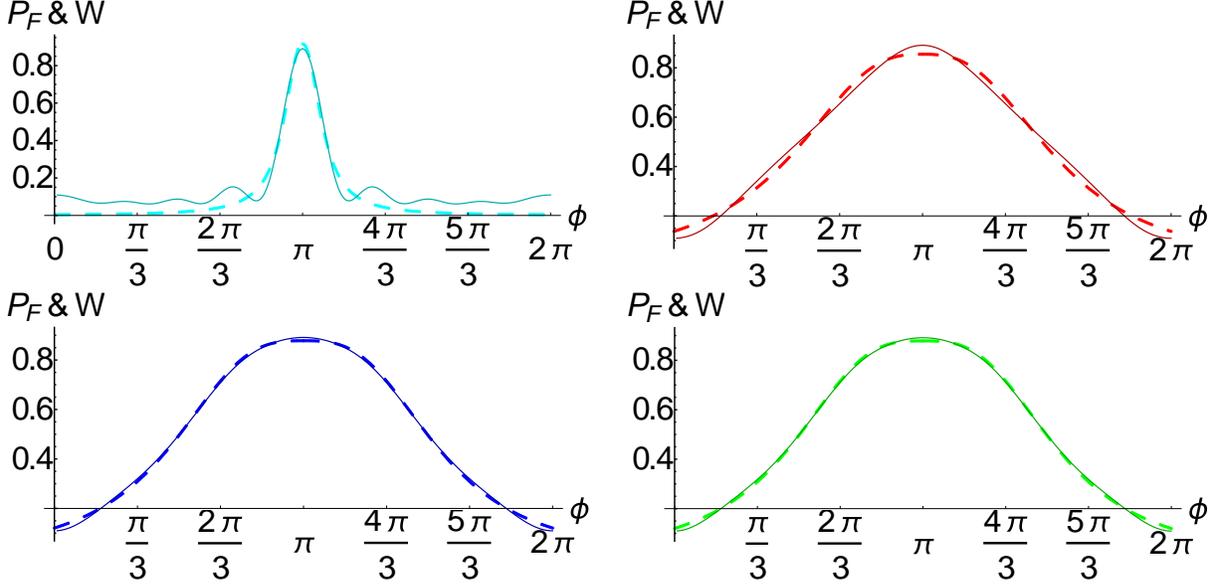

*Fig. 10. Comparison between the Hill fit and the simulation of QRWS with first neighbours included when the coin size is 6. With solid dark lines are the probabilities to find solution for functional dependence given by Eq. (20) (top left), Eq. (21) (top right), Eq. (22) (bottom left) and Eq. (23) (bottom right). The dashed lines correspond to Hill fits for the same equations.*

Similarly, to [19], we make approximations with Hill functions of the QRWS algorithm with coin sizes between 4 and 10, for each functional dependence $\zeta(\phi)$. Next, the Hill function's parameters $b, k$, and $n$ for each $\zeta(\phi)$ and all simulated coin sizes m, are fitted a second time to obtain the functions $b_i(m)$, $k_i(m)$ and $n_i(m)$, where the index i runs through all $\zeta(\phi)$.

We will use the same fitting functions as in [19]. However, for $k_i(m)$ a generalization will be made that has better performance for all simulations data, including the unmodified QRWS, inclusion of first and first and second neighbours:

$$b_i(m) = c_1/x + c_2 \quad (30)$$

$$k_i(m) = c_1 \cdot e^{c_2 x} \cdot x^{c_3} + c_4 \quad (31)$$

$$n_i(m) = c_1 \cdot x^2 + c_2 \cdot x + c_3 \quad (32)$$

The results for parameter's values obtained by the fit with those functions is shown on Table. 2:

| Neighbors | Functional Dependence | Parameter | Secondary Fit Parameters | | | |
|---|---|---|---|---|---|---|
| | | | $c_1$ | $c_2$ | $c_3$ | $c_4$ |
| None | $\zeta = const$ | $b_1$ | -0.285537 | 0.452847 | ------------ | ------------ |
| | | $k_1$ | 1.07753 | -0.446902 | 0.720219 | 0 |
| | | $n_1$ | 9.3266 $10^{-6}$ | 0.00802087 | 3.19524 | ------------ |
| | $\alpha = 0$ | $b_2$ | -0.410465 | 0.4704 | ------------ | ------------ |
| | | $k_2$ | 1.12984 | -0.513774 | 1.97337 | 0 |
| | | $n_2$ | 0.1162 | -2.08982 | 11.4129 | ------------ |
| | $\alpha = -\dfrac{1}{2\pi}$ | $b_3$ | −0.289659 | 0.451867 | ------------ | ------------ |
| | | $k_3$ | 6.33443 | 0 | -0.761267 | 0.120075 |
| | | $n_3$ | 0.250955 | −3.19373 | 13.4087 | ------------ |
| | $\alpha = \alpha_{ML}$ | $b_4$ | −0.288605 | 0.451698 | ------------ | ------------ |
| | | $k_4$ | 6.03959 | 0 | -0.675635 | -0.0525597 |
| | | $n_4$ | 0.193981 | −2.33236 | 10.6187 | ------------ |
| First | $\zeta = const$ | $b_1$ | -0.468471 | 0.968085 | ------------ | ------------ |
| | | $k_1$ | 1.07763 | -0.531884 | 1 | 0 |
| | | $n_1$ | 0.0418295 | -0.468643 | 3.56198 | ------------ |
| | $\alpha = 0$ | $b_2$ | -0.69904 | 0.983061 | ------------ | ------------ |
| | | $k_2$ | 2.91787 | -0.381125 | 1 | 0 |
| | | $n_2$ | 0.0242485 | -0.642224 | 5.72698 | ------------ |
| | $\alpha = -\dfrac{1}{2\pi}$ | $b_3$ | -0.343152 | 0.932635 | ------------ | ------------ |
| | | $k_3$ | 13.2362 | 0 | -1.41297 | 0.695604 |
| | | $n_3$ | 0.178899 | -2.04328 | 8.80694 | ------------ |
| | $\alpha = \alpha_{ML}$ | $b_4$ | -0.352484 | 0.934448 | ------------ | ------------ |
| | | $k_4$ | 12.938 | 0 | -1.39904 | 0.699605 |
| | | $n_4$ | 0.15006 | -1.55603 | 7.1642 | ------------ |
| Second | $\zeta = const$ | $b_1$ | -0.776623 | 1.18983 | ------------ | ------------ |
| | | $k_1$ | 4.61967 | -0.741446 | 1 | 0 |
| | | $n_1$ | 0.101875 | -1.32993 | 5.12931 | ------------ |
| | $\alpha = 0$ | $b_2$ | -0.449375 | 1.04878 | ------------ | ------------ |
| | | $k_2$ | 274.205 | 0.123303 | -3.19942 | 0.10206 |
| | | $n_2$ | 7.7711 $10^{-6}$ | 0.020196 | 1.53204 | ------------ |
| | $\alpha = -\dfrac{1}{2\pi}$ | $b_3$ | -0.076026 | 0.989097 | ------------ | ------------ |
| | | $k_3$ | 1027.74 | 0 | -4.02775 | 1.1494 |
| | | $n_3$ | 0.122989 | -1.10132 | 4.12241 | ------------ |
| | $\alpha = \alpha_{ML}$ | $b_4$ | -0.0355965 | 0.981581 | ------------ | ------------ |
| | | $k_4$ | 1037.96 | 0 | -4.02422 | 1.138691 |
| | | $n_4$ | 0.122229 | -1.00595 | 3.6904 | ------------ |

*Table. 2. The fitting parameters that gives the best Hill secondary fitting of probability to find solution in three cases: 1) For marked state only $P_W(m, \phi)$; 2) Marked state together with its first neighbors $P_F(m, \phi)$ 3) Marked state together with its first and second neighbors $P_W(m, \phi)$. Those calculations are made for different functional dependences between the phases.*

### 4.2.3. Robustness when the first and both the first and second neighbors are included

The Hill function (29), together with the fitting functions of its parameters given in Table. 2. can be used to make extrapolation of the robustness when the coin size is larger than 10. As validation we can use the results for coin size 11 that can be both extrapolated and obtained by numerical simulation of the algorithm.

On Fig. 11 are shown the results for the robustness of QRWS (defined in Sec. 3.3.) without any neighbors $\varepsilon_W$ (top left), with the first neighbors $\varepsilon_F$ (top right) and with the first and second neighbors $\varepsilon_S$ (bottom). The numerical simulation of the algorithm is used for coin sizes between 4 and 11. The robustness for coin sizes between 11 and 25 is obtained by using extrapolations based on Hill function fits.

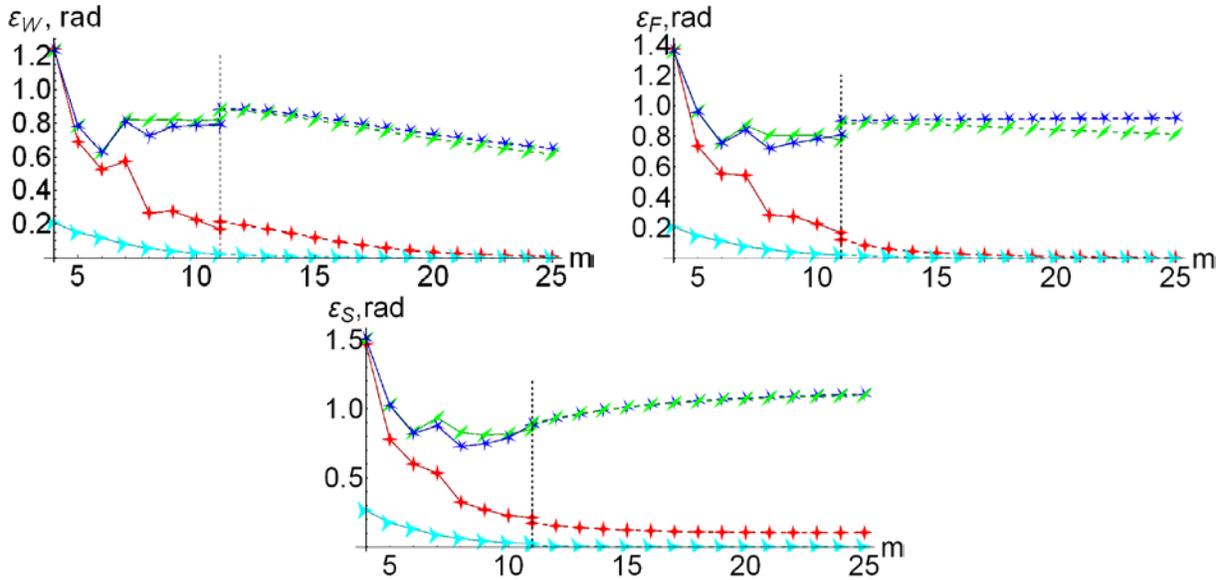

*Fig. 11. Robustness of QRWS algorithm on hypercube calculated by numerical simulation for coin sizes between 4 and 11 (depicted by straight lines). For coin sizes between 11 and 25 a prognosis was made based on Hill function fits (depicted by dashed line). The top left figure corresponds to the QRWS without taking additional neighbours. On the top right is the robustness of QRWS when in addition to the marked state, its first neighbours are taken. Similarly, on the bottom is the robustness when together with the marked state, both first and second neighbours are taken. Different colour and marker correspond to different dependence $\zeta(\phi)$. The teal three pointed star is used for Eq. (20), the red four pointed star to Eq. (21), the blue five pointed star to Eq. (22) and the green two pointed star to Eq. (23).*

Similarly to QRWS without including the nearest neighbors, using nonlinear dependences (Eq. (22) and Eq. (23)) gives much better results than the linear ones (Eq. (20) and Eq. (21)) when marked state's neighbors are included.

The robustness for small coin sizes in case of first neighbors and Eq. (21) is initially high but decreases fast with increasing the coin size and we prognose that for coin sizes around 15 it will become close to the one of Eq. (20). Our model shows that adding second neighbors gives small change in the robustness for small coin size. However, for large coin size the robustness of QRWS when phases are related by Eq. (20) goes to fixed value.

The values for the robustness when the dependences between the phases are according to Eq. (22) and Eq. (23) are pretty close with and without inclusion of both first and second neighbors. It can be seen that in the case of inclusion of the first neighbors $\alpha = -1/2\pi$ gives higher robustness than $\alpha = \alpha_{ML}$. The reason behind this is that the neural network that was used in [18] to obtain $\alpha_{ML}$ was trained on data when there were no neighbors taken. We assume that the values of $\alpha_{ML}$ when the first neighbors are taken are different than the ones obtained from simulations of the unmodified QRWS.

For dimensions between 4 and 11 the robustness of QRWS algorithm is higher with inclusion of first neighbors (when the functional dependence between phases is the same). However, with increasing the coin size the difference becomes smaller.

In the case of second neighbor, the robustness with both Eq. (22) and Eq. (23) increases while increasing the coin size for coin dimension between 4 and 11. Our prognosis shows that the robustness will increase even more for larger coin size.

An alternative method to investigate the robustness of QRWS when first and second neighbors are included is shown in Appendix 2. Results obtained in this way complement the results in this section. However, the method in the appendix cannot be used for extrapolations for larger coin sizes.

In the next section we will make prognosis how robustness changes for large coin sizes by using the explicit form of the fitting functions found, and set m to infinity.

### 4.2.4. Robustness comparison for large coin size

By using the Hill function, we can make an estimation how the robustness changes for large coin size and for each of the Eqs. (20-23) when only the first $\varepsilon_F/\varepsilon_W$ and when both first and second $\varepsilon_S/\varepsilon_F$ neighbors are included. We define the robustness $\tilde{\varepsilon}$ derived from the Hill function fits as:

$$W(\pi + \tilde{\varepsilon}, m) = \Omega\, W(\pi, m) \tag{33}$$

Then, we have:

$$\tilde{\varepsilon} = \left(\frac{(1-\Omega)k_i(m,\zeta(\phi))^{n_i(m,\zeta(\phi))}}{\Omega}\right)^{\frac{1}{n_i[m]}} \tag{34}$$

And the robustness relations are:

$$\frac{\varepsilon_F}{\varepsilon_W} \cong \frac{\tilde{\varepsilon}_F}{\tilde{\varepsilon}_W} = \left(\frac{1-\Omega}{\Omega}\right)^{\frac{n_{i,W}(m,\zeta(\phi))-n_{i,F}(m,\zeta(\phi))}{n_{i,F}(m,\zeta(\phi))n_{i,W}(m,\zeta(\phi))}} \frac{k_{i,F}(m,\zeta(\phi))}{k_{i,W}(m,\zeta(\phi))} \tag{35}$$

$$\frac{\varepsilon_S}{\varepsilon_F} \cong \frac{\tilde{\varepsilon}_S}{\tilde{\varepsilon}_F} = \left(\frac{1-\Omega}{\Omega}\right)^{\frac{n_{i,F}(m,\zeta(\phi))-n_{i,S}(m,\zeta(\phi))}{n_{i,F}(m,\zeta(\phi))n_{i,S}(m,\zeta(\phi))}} \frac{k_{i,S}(m,\zeta(\phi))}{k_{i,F}(m,\zeta(\phi))} \tag{36}$$

In case of Eq. (23) and Eq. (22) there is a plateau at the maximum of the fit with Hill function, so the width of $W(\phi, m)$ is primary defined by $k_{i,F}(m,\zeta(\phi)) = c_1 e^{c_2} m^{c_3} + c_4$. In this case $W(\phi, m)$ has shape close to a square pulse, so it should depend very weakly on $\Omega$. Then the term containing $\Omega$ and n could be neglected. Now, the relations between the robustness of the three cases simplifies to:

$$\frac{\varepsilon_F}{\varepsilon_W} \cong \frac{k_{i,F}(m,\zeta(\phi))}{k_{i,W}(m,\zeta(\phi))} \cong \frac{c_{1,k_i,F} \cdot e^{c_{2,k_i,F} x} \cdot m^{c_{3,k_i,F}} + c_{4,k_i,F}}{c_{1,k_i,W} \cdot e^{c_{2,k_i,W} x} \cdot m^{c_{3,k_i,W}} + c_{4,k_i,W}} \tag{37}$$

$$\frac{\varepsilon_S}{\varepsilon_F} \cong \frac{k_{i,S}(m,\zeta(\phi))}{k_{i,F}(m,\zeta(\phi))} \cong \frac{c_{1,k_i,S} \cdot e^{c_{2,k_i,S} x} \cdot m^{c_{3,k_i,S}} + c_{4,k_i,S}}{c_{1,k_i,F} \cdot e^{c_{2,k_i,F} x} \cdot m^{c_{3,k_i,F}} + c_{4,k_i,F}} \tag{38}$$

In case of Eq. (21) and Eq. (20) there is not such plateau and functions' shapes are not similar to the ones of the square pulse. In these cases, the formula that estimates how the robustness changes depending on the number of neighbors cannot be significantly simplified, due to the impossibility of taking a sufficiently accurate values of the fitting parameters, leading to large approximation errors.

## 5. Conclusion

In this work we study the robustness of random walk search algorithm in case when we take the nearest neighbors of the measured state. Our previous works show that modification of quantum random walk search algorithm with traversing coin constructed by generalized Householder reflection and phase multiplier with particular dependence between the coin phases can makes the algorithm more robust (insensitive against inaccuracies in the phases). It is already known that measuring both the state that the algorithm returns together with its first neighbors gives more than two-fold increase in the probability to find solution. Using each of those modifications improves the QRWS algorithm. Here, we study the possibility to combine them together.

We show that using particular functional dependences between the phases increases the robustness of the algorithm when no additional neighbors are taken. When the first neighbors of the marked state are included, a similar increase in stability is observed. Adding the first neighbors increases the robustness of the algorithm on the price of slightly increasing the needed number of measurements. Taking its second neighbors requires even more measurements, however the Hill function extrapolations predict consistently high robustness of the QRWS algorithm for large coin size.

We compare the robustness of the three cases (no neighbors, first neighbors and second neighbors) for coin size between 4 and 11 and use an approximation by Hill function to make a prediction for larger con sizes. We show that for fixed phase – the percentage of their maximal probability they achieve slightly differs in cases of no neighbors, only 1-st and both 1-st and 2-nd. This depends on both the value of the angle in generalized Householder reflection and the functional relation between both coin angles.

### Acknowledgments

The work on this paper was supported by the Bulgarian National Science Fund under Grant KP-06-N58/5 / 19.11.2021.

## Appendix 1: Robustness of QRWS with the first and second neighbors included for more coin sizes

Here, on Fig. 15 we show results of numerical simulations of the probability to find solution of QRWS algorithm in cases of including: only the first neighbors $P_F(\phi, \zeta(\phi), m)$ (on the left column) or both the first and second neighbours $P_S(\phi, \zeta(\phi), m)$ (on the right column). Different colour and dashing correspond to different functional dependence between the phases $\zeta(\phi)$. On the figure, with red dot-dashed, teal dashed, solid green and dotted blue lines are shown results for dependences according to Eq. (21), Eq. (20), Eq. (23), Eq. (22) correspondingly. Each row corresponds to different coin size.

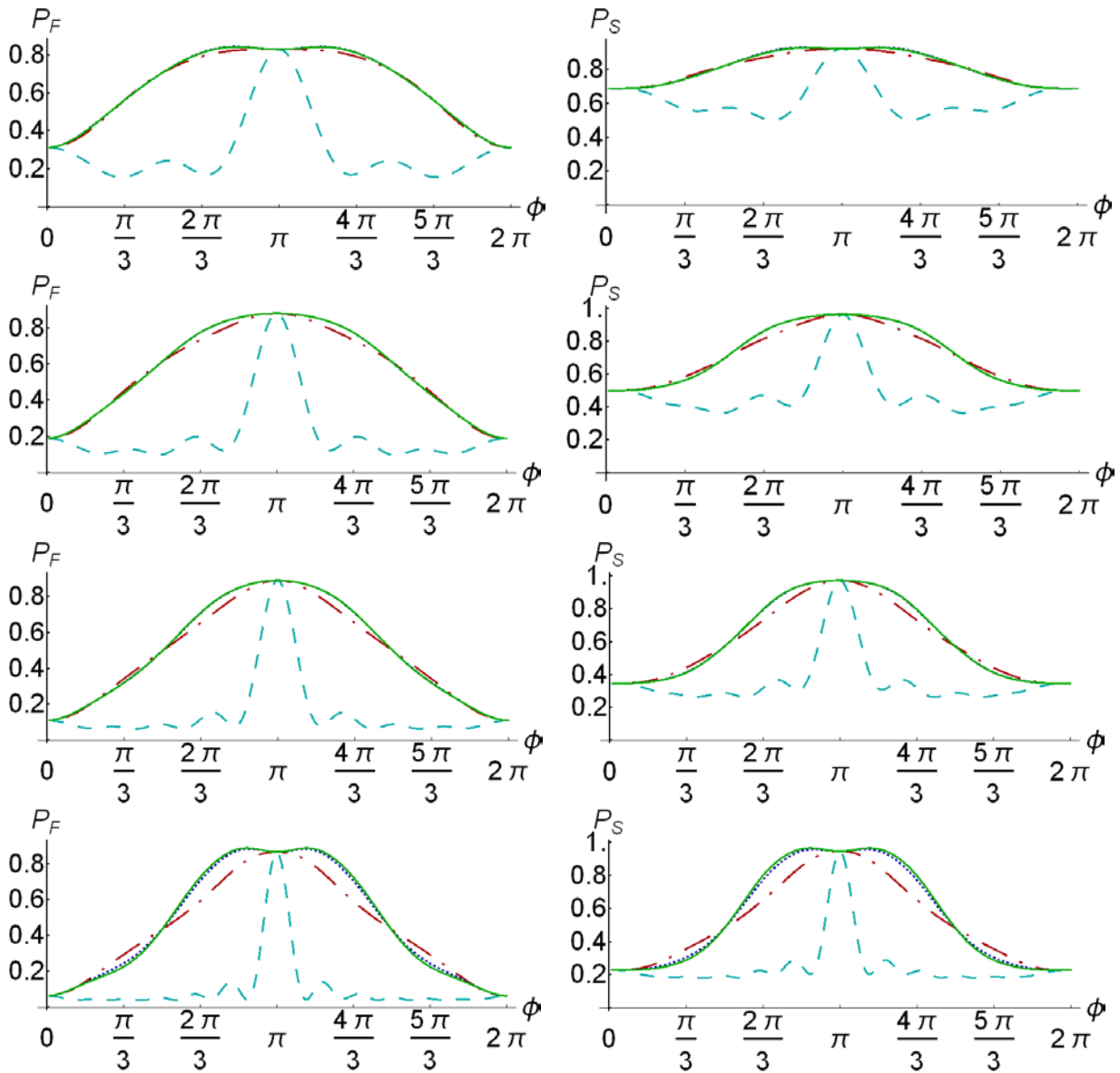

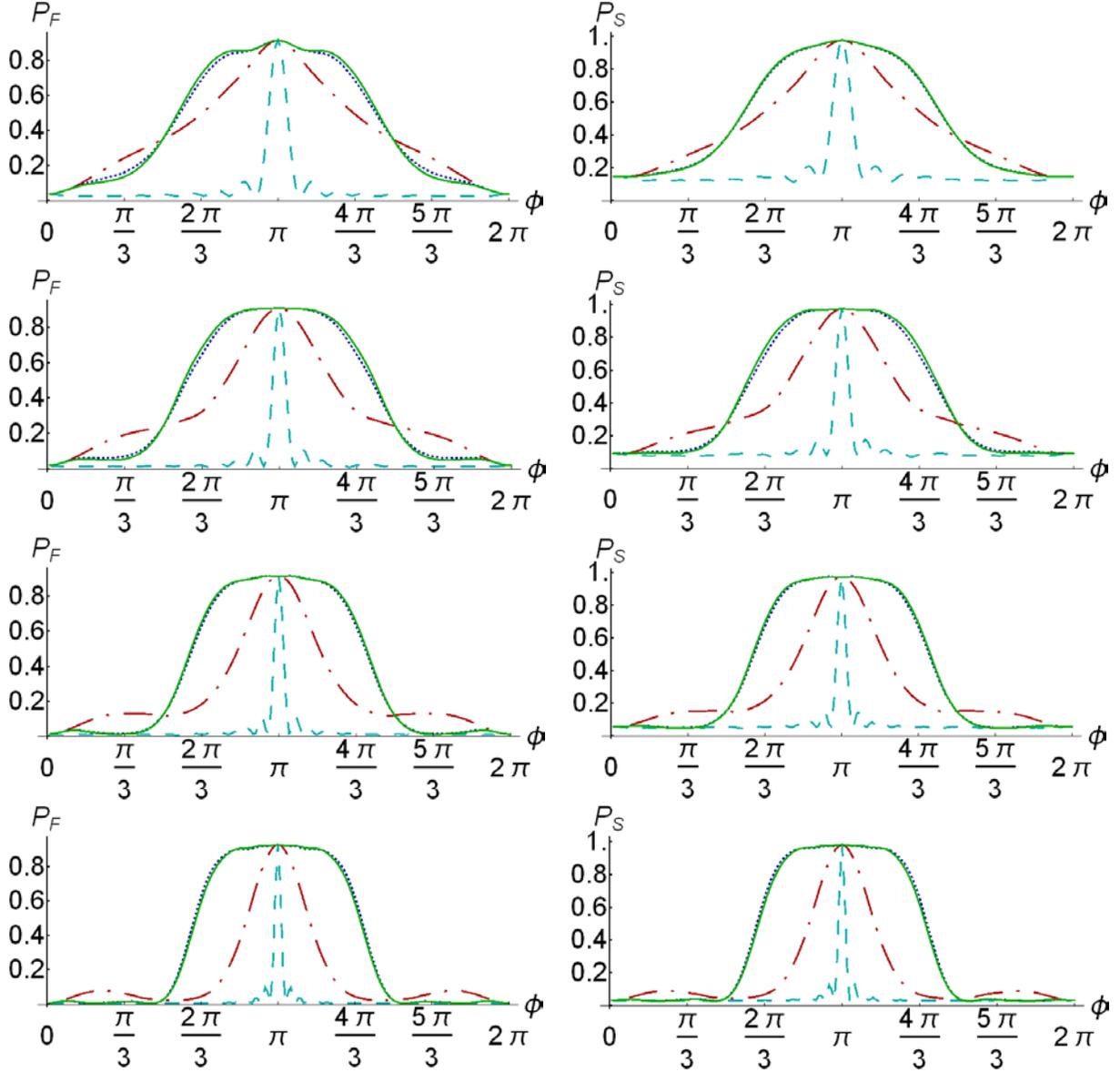

*Fig. 12. Probability to find solution of QRWS when only first neighbours $P_F(\phi, \zeta(\phi), m)$ (on left) and both first and second neighbours $P_S(\phi, \zeta(\phi), m)$ (on right) are included. Each row corresponds to different coin size: m=4 on first, m=5 on second and so on until the eight row, where m=11. Different color and dashing correspond to different functional dependence between the phases: the red dot-dashed, the teal dashed, the solid green and the dotted blue lines correspond to Eq. (21), Eq. (20), Eq. (23), Eq. (22) respectively.*

## Appendix 2: Comparison in fixed interval

Sometimes it is important to compare the stability of functions in a fixed finite interval. To be able to compare them, we first define new variables (Eq.(45)): the first one is the ratio between the normalized probability to find solution when the first neighbors are taken $P_F$ divided by the normalized probability without any neighbors $P_W$. Similarly, we define the ratio of the normalized success rate when both first and second neighbors are measured $P_S$ and the normalized probability when only the first neighbors are measured:

$$\lambda_1(m,\phi,\zeta(\phi)) = \left(\frac{P_F(m,\phi,\zeta(\phi))}{P_F(m,\pi,\zeta(\pi))}\right) \cdot \left(\frac{P_W(m,\phi,\zeta(\phi))}{P_W(m,\pi,\zeta(\pi))}\right)^{-1} \quad (39)$$

$$\lambda_2(m,\phi,\zeta(\phi)) = \left(\frac{P_S(m,\phi,\zeta(\phi))}{P_S(m,\phi,\zeta(\phi))}\right) \cdot \left(\frac{P_F(m,\phi,\zeta(\phi))}{P_F(m,\pi,\zeta(\pi))}\right)^{-1} \quad (40)$$

The values $\lambda_1(m,\phi,\zeta(\phi))$ and $\lambda_2(m,\phi,\zeta(\phi))$ shows which of those functions ($P_W$, $P_W$, and $P_s$) is closer to its maximal value in a fixed point $\phi$. If there is area around $\phi = \pi$ where $\lambda_1(m,\phi,\zeta(\phi)) = 1$ then in this area both unmodified QRWS and the modification that takes the first neighbours are equally close to its maximal value. If there are area where $\lambda_1(m,\phi,\zeta(\phi)) > 1$ then the modification that takes first neighbors is closer to its peak, and when $\lambda_1(m,\phi,\zeta(\phi)) < 1$ then the unmodified QRWS on hypercube shows better behavior. Similarly, $\lambda_2(m,\phi,\zeta(\phi))$ shows how much taking first or second neighbors improve or worsen the stability in the interval.

On Fig. 12 are shown $\lambda_1(m,\phi,\zeta(\phi))$ and $\lambda_2(m,\phi,\zeta(\phi))$ in case of coin sizes $m=6$ (top and bottom left correspondingly) and for $m=10$ (top and bottom right correspondingly). Different functional dependences Eq. (21), Eq. (22), Eq. (23) corresponds to dot-dashed red, dotted blue and solid green curves. The resulted curve for Eq. (20) is excluded because $\lambda(m,\phi,\zeta(\phi))$ gives too high numerical errors due to divisions of very small numbers.

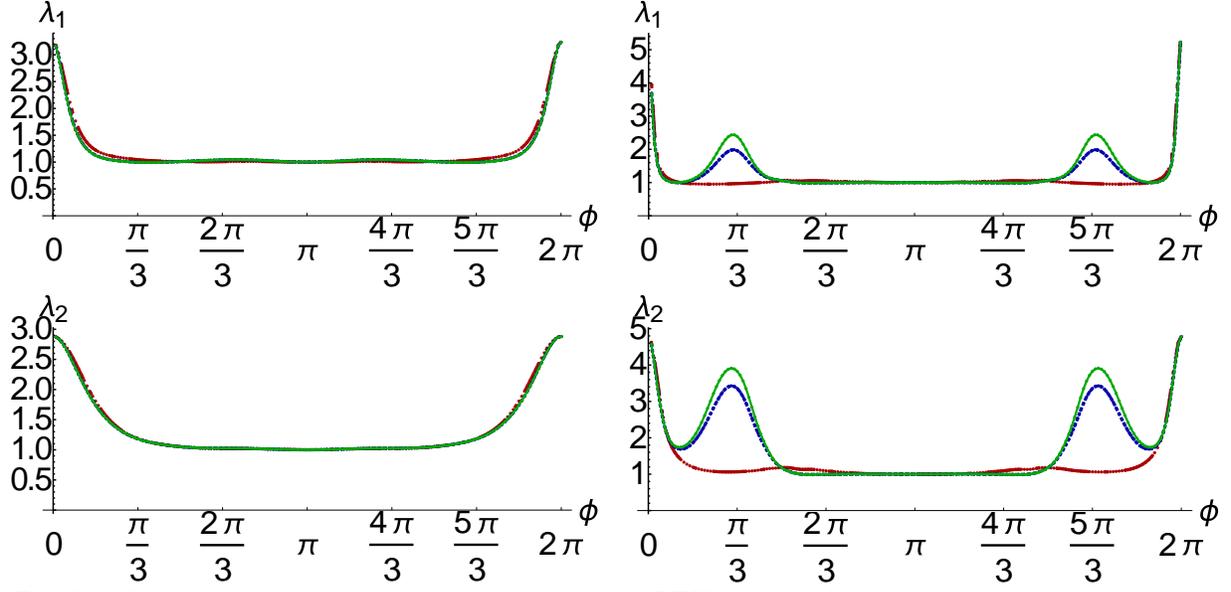

*Fig. 13. Comparison between robustness of the QRWS with and without including neighbours as is defined by Eq. (45) and Eq. (46). Dash-dotted red, doted blue and solid green lines corresponds to Eq. (21), Eq. (22) and Eq. (23). The robustness is compared in the interval between 0 and 2π. Left pictures correspond to m=6, and right to m=10*

As expected, when the coin is with $\phi \gtrsim 0$ or $\phi \lesssim 2\pi$ the slope of $\lambda_1$ and $\lambda_2$ increases with increasing the coin size. This is due to big differences close to those points between $P_F(m, \phi, \zeta(\phi))$, $P_S(m, \phi, \zeta(\phi))$ and $P_W(m, \phi, \zeta(\phi))$.

To see more clearly the result for a particular interval of interest we need to compare results for all points in the interval. Another reason is that at the points $\phi = 0$ and $\phi = \pi$, the value of $\lambda$ is much higher than at the points in the center (which are more important for our needs). That's why we need to take interval far from $\phi = 0$ and $\phi = \pi$.

On Fig. 13 we show $\lambda_1$ (first row) and $\lambda_2$ (second row) for coin size 6 (first column) and 10 (second column) in interval between $2\pi/3$ and $4\pi/3$. Red dash-doted, dotted blue and solid green lines corresponds to Eq. (21), Eq. (22), Eq. (23) respectively.

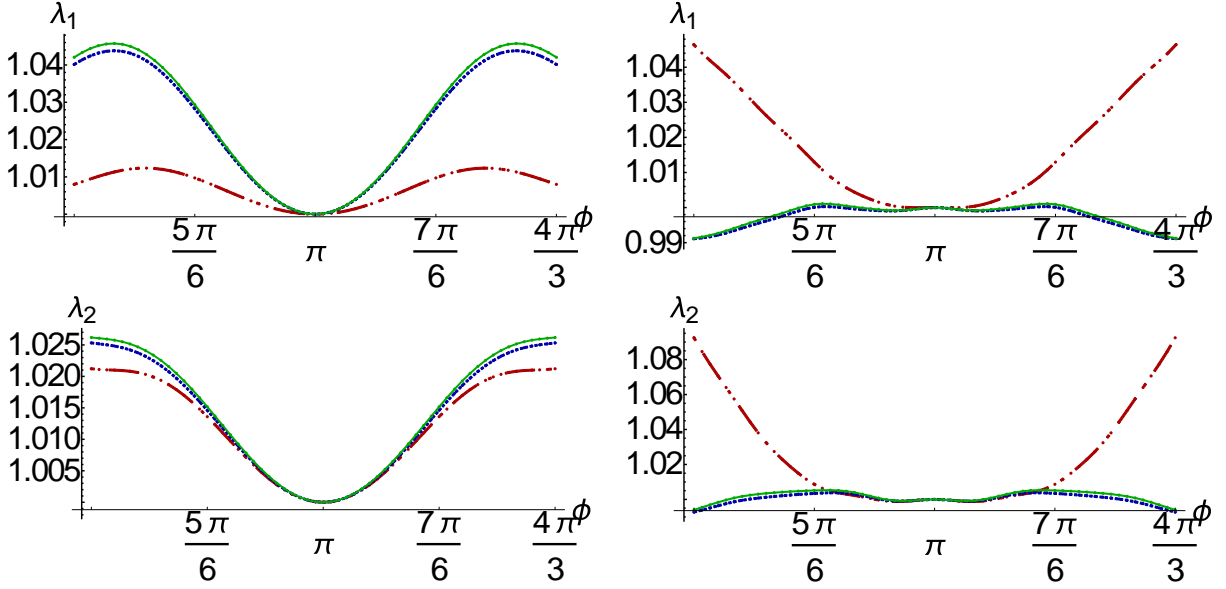

*Fig. 14. Comparison between robustness of the QRWS with and without including neighbours as is defined by Eq. (45) (first row) and Eq. (46) (second row). The dash-dotted red, doted blue and solid green lines corresponds to Eq. (21), Eq. (22) and Eq. (23). The robustness is compared in the interval between $2\pi/3$ and $4\pi/3$. Left pictures correspond to m=6, and right to m=10*

In the interval $\phi \in [2\pi/3, 4\pi/3]$, we can see how taking additional neighbors increases $\lambda_1(m, \phi, \zeta(\phi))$ and $\lambda_2(m, \phi, \zeta(\phi))$. Their values depend on the number of neighbors taken, the coin size and the functional dependence between phases. However, the values of $\lambda_1$ and $\lambda_2$ on particular values of $\phi$ can be more or less than one and the visual comparison is not always reliable. We need to compare the robustness in a fixed interval like between $2\pi/3$ and $4\pi/3$ (both values above are taken as an illustrative example). Now we will continue our analysis for the interval with higher importance $(\pi - \varepsilon, \pi + \varepsilon)$. Additional complications come from the fact that $\varepsilon$ depends on both the coin size m and the functional dependence between the phases $\zeta(\phi)$. To have a good measure for the robustness we need to calculate its average for the particular coin size and functional dependence in the interval were the standard QRWS is robust:

$$\Lambda_1(m, \zeta) = \left( \frac{\int_\pi^{\pi+\varepsilon} \lambda_1(m, \phi, \zeta(\phi)) d\phi}{\varepsilon(m, \zeta)} - 1 \right) \quad (41)$$

$$\Lambda_2(m, \zeta) = \left( \frac{\int_\pi^{\pi+\varepsilon} \lambda_2(m, \phi, \zeta(\phi)) d\phi}{\varepsilon(m, \zeta)} - 1 \right) \quad (42)$$

We need to integrate and divide only on the half of the interval because due to symmetry of the function $p(\phi, \zeta(\phi))$ about the point $\phi = \pi$, the functions $\lambda_1(m, \phi, \zeta(\phi))$, $\lambda_2(m, \phi, \zeta(\phi))$, $\Lambda_1(m, \zeta)$ and $\Lambda_2(m, \zeta)$ are also symmetric. The results for numerical calculations for $\Lambda_1(m, \zeta)$ and $\Lambda_2(m, \zeta)$ in the case of coin size between 4 and 11 are shown in Table. 5:

| m | $\Lambda_1(m, \zeta)$ | | | $\Lambda_2(m, \zeta)$ | | |
|---|---|---|---|---|---|---|
|  | Eq. (21) | Eq. (22) | Eq. (23) | Eq. (21) | Eq. (22) | Eq. (23) |
| 4 | 0.005802 | 0.01951 | 0.01755 | −0.008977 | −0.006607 | −0.007307 |
| 5 | 0.003309 | 0.01520 | 0.01547 | 0.002585 | 0.004269 | 0.004389 |
| 6 | 0.003793 | 0.01406 | 0.01460 | 0.005145 | 0.007295 | 0.007573 |
| 7 | −0.004817 | 0.005987 | 0.01028 | −0.001588 | −0.004873 | −0.002370 |
| 8 | 0.004189 | 0.007603 | 0.007941 | 0.008235 | 0.01582 | 0.01791 |
| 9 | −0.001113 | −0.004875 | −0.004193 | −0.001774 | −0.0015832 | 0.0008598 |
| 10 | 0.0002006 | −0.001024 | −0.0006164 | −0.0007097 | 0.001567 | 0.002702 |
| 11 | -0.0003082 | 0.0051 | 0.004581 | 0.001469 | 0.005268 | 0.004021 |

*Table. 3. Numerical calculations for $\Lambda_1(m, \zeta)$ and $\Lambda_2(m, \zeta)$. Each coin size $m \in [4,11]$ is shown on different row and the columns corresponds to different functional dependence between the angles.*

The Fig. 14 shows absolute values of $\Lambda_1(m, \zeta)$ and $\Lambda_2(m, \zeta)$ for coin sizes between 4 and 11. Different functional dependences Eq. (21), Eq. (22), Eq. (23) corresponds to dot-dashed red, dotted blue and solid green curves.

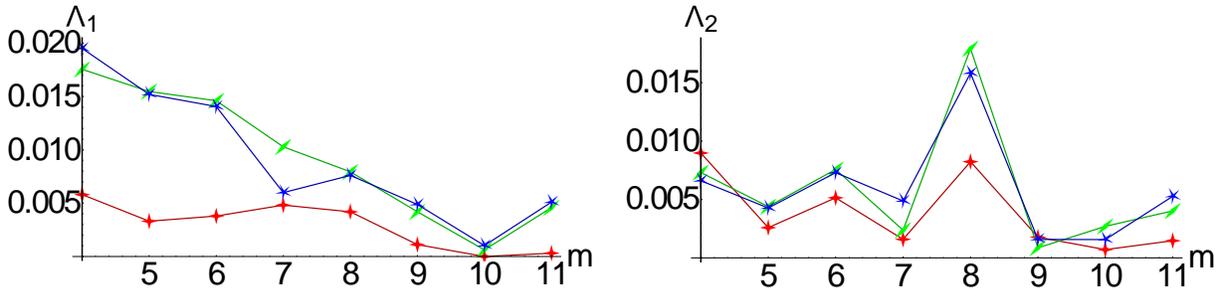

*Fig. 15. Comparison of the numerical calculations for $|\Lambda_1(m, \zeta)|$ and $|\Lambda_2(m, \zeta)|$. Each functional dependence is shown on different color and different star. Red line with four point star, blue line with five point star and green line with two point star corresponds to Eq. (21), Eq. (22) and Eq. (23) respectively.*

Adding the first neighbors for particular value of the coin size sometimes increases and sometimes decreases the robustness. Similar is the result when the second neighbors are added. With increase of the coin size the absolute value of the additional robustness that comes from the first neighbors decreases. Adding second neighbors changes the absolute value of the robustness even less. The value of $\Lambda_1(m,\zeta)$ decreases with the coin size, and $\Lambda_2(m,\zeta)$ have too chaotic behavior to be able to make prognosis how it changes with increasing the coin size.

The subtle details of $P_S(m,\phi)$, $P_F(m,\phi)$ and $P_W(m,\phi)$ cannot be guessed by using the extrapolation with Hill function for larger coin size. However, the overall trend is that even for large coins adding more neighbors increases the robustness.